\documentclass[useAMS,usenatbib]{mn2e}
\usepackage{graphicx}
\def\iint{\int\!\!\!\!\!\int}
\def\iiint{\iint\!\!\!\!\!\int}
\usepackage{scrextend}
\title[Gas distribution in MS 1054-0321]{The gas distribution in the
  high-redshift cluster MS 1054-0321}
\author[M. S. Mirakhor and M. Birkinshaw]{M. S. Mirakhor\thanks{E-mail:
mohammad.mirakhor@bristol.ac.uk} and M. Birkinshaw\\H. H. Wills
Physics Laboratory,
University of Bristol, Tyndall
Avenue, Bristol BS8 1TL}

\begin{document}

\date{Accepted 2016 January 18. Received 2016 January 15; in original form 2015 October 06}

\pagerange{\pageref{firstpage}--\pageref{lastpage}} \pubyear{2016}

\maketitle

\label{firstpage}

\begin{abstract}
We investigate the gas mass distribution in the
high redshift cluster MS 1054-0321 using {\it Chandra \/} X-ray
and OCRA SZ effect data. We use  a superposition of offset $\beta$-type
models to describe the composite structure of MS 1054-0321. We find
gas mass fractions
$f_{gas}^\rmn{X\mbox{-}ray}=0.087_{-0.001}^{+0.005}$ and
$f_{gas}^\rmn{SZ}=0.094_{-0.001}^{+0.003}$ for the (main) eastern
component of MS 1054-0321 using X-ray or SZ data, but
$f_{gas}^\rmn{X\mbox{-}ray}=0.030_{-0.014}^{+0.010}$ for the
western component. The gas mass fraction for the eastern component is
in agreement with some results reported in the literature, but
inconsistent with the cosmic baryon fraction. The low gas mass
fraction for the western component is likely to be a consequence of
gas stripping during the ongoing merger. The gas mass fraction of the
integrated system is $0.060_{-0.009}^{+0.004}$: we suggest that the
missing baryons from the western component are present as hot
diffuse gas which is poorly represented in existing X-ray images. The
missing gas could appear in sensitive SZ maps.


\end{abstract}

\begin{keywords}
galaxies: clusters: individual: MS 1054-0321 $-$ galaxies: clusters:
intracluster medium $-$ galaxies: clusters: general $-$ X-rays:
galaxies: clusters $-$ cosmic background radiation $-$ cosmology: observations
\end{keywords}

\section{Introduction}
The properties of massive high redshift clusters could place powerful
constraints on the cosmological parameters that govern the physics of
the early Universe \citep{b12,b4}. Several studies have used  clusters
at high redshift to provide valuable information on
galaxy formation and evolution in the richest environments and to
probe Gaussianity in the density field that seeded structure formation
\citep{b9,b20,b4}. Massive clusters, with $M_{tot}\sim 10^{15}\,\rmn{M_\odot}$, are
identified by their gravitational lensing of background galaxies \citep{b41,b42},
and by the X-ray emission or Sunyaev-Zel'dovich effect (SZ) from their
hot intracluster medium (ICM) \citep{b43,b44}.

The X-ray emission from diffuse hot plasma at $ T > 5 \,\rmn{keV}$ is primarily
bremsstrahlung from the gas within the potential well of the cluster's
dark matter halo. The X-ray luminosity is proportional to the square
of the electron density, hence is sensitive to the denser regions of
the cluster. On the other hand, the thermal SZ effect, the change in
the brightness of the cosmic microwave background caused by hot electrons in
the ICM inverse-Compton scattering, depends on the ICM pressure (the
integrated product of the electron density and temperature on the line
of sight). This parameter dependence makes the SZ effect a
  powerful tool to probe lower-density regions of the hot gas, and
  thus to provide better insight into gas properties outside the
cluster core or in the hottest regions. For reviews of the physics of
the SZ effect see \citet{b18}, \citet{b2}, and
\citet{b5}. Due to the different sensitivity of the SZ effect and the X-ray
emission to outer and inner regions of the cluster, these two
independent probes can be used to provide a more complete picture of the
ICM. By combining X-ray and SZ effect data, the distance of
the cluster can be estimated and hence the Hubble constant or other
cosmological parameters can be found
\citep[e.g.][]{b1,b21,b3}. Alternatively, tight constraints on the gas
mass fraction of the cluster can be obtained \citep[see][for
example]{b11,b14,b16}.

Observations with X-ray satellites such as {\it Chandra\/} and {\it XMM--Newton\/}
have provided high quality images of the X-ray emission and
detailed structural properties of the ICM \citep[e.g.][]{b19,b17}. In
parallel, many SZ observatories such as the South Pole Telescope
\citep[SPT,][]{b6}, and the One Centimeter Receiver array
\citep[OCRA,][]{b13} have made high signal-to-noise SZ measurements
toward clusters. Generally the models used to interpret SZ effect have
been rather simple.

In this paper we investigate the gas mass distribution of the high
redshift cluster MS 1054-0321, taking advantage of a long {\it Chandra\/}
observation of MS 1054-0321 and high signal/noise SZ effect data from
OCRA, to develop a three-dimensional gas density profile of the
$\beta$-model type. We use this to improve the measurement of the gas
mass of MS 1054-0321. The physical properties of the gas of MS 1054-0321
were analysed in previous studies. However in most of these studies, standard isothermal
$\beta$-models were used for the cluster plasma distribution although
the structure of the cluster is clearly more complicated. Since the
ICM mass and the total mass of cluster are functions of temperature,
we also re-analyse the X-ray temperature structure of MS
1054-0321. We use our results to estimate the gas and total masses of
the cluster.

In the next Section we review MS 1054-0321. In Section 3
we describe the X-ray and radio observations of MS 1054-0321. The
analysis method for 3D single-$\beta$ and double-$\beta$ models
used to describe the plasma distribution is presented in Section
4. The gas mass content of MS 1054-0321 is discussed in Section
5. Finally, in Section 6 we summarise our results. Throughout this
paper, we use a $\Lambda$CDM cosmology with $\Omega_{\rmn{m}} = 0.3$,
$\Omega_{\Lambda} = 0.7$, and $H_0 = 100 \,h_{100} \,\rmn
{km\,sec^{-1}\,Mpc^{-1}}$ with $h_{100}=0.7$. Uncertainties are at the
68\% confidence level, unless otherwise stated.

\section{CLUSTER MS 1054-0321}

The rich cluster MS 1054-0321 is the most distant and
luminous cluster in the Einstein Extended
Medium Sensitivity Survey \citep[EMSS,][]{b10}. Due to its high
redshift  ($z$ = 0.83 corresponding to angular diameter distance $D_A=
1.57\,\rmn{Gpc}$) and complicated morphology, the ICM plasma of
MS 1054-0321 has been subjected to many studies. X-ray observations of
MS 1054-0321 with the {\it ASCA\/} and {\it ROSAT\/} satellites were
analysed by \citet{b7}. The {\it ASCA\/} spectrum showed a high X-ray
temperature of $12.3^{+3.1}_{-2.2} \,\rmn{keV}$ ($90\%$ confidence
interval), which should indicate that MS 1054-0321 is a massive
cluster. The X-ray image from the {\it ROSAT HRI\/} identified
two or three components in addition to an extended component,
suggesting that MS 1054-0321 is not in a relaxed state. Based on this
X-ray temperature and using the mass-temperature relation adopted from
the simulations of \citet{b8}, the virial mass of MS 1054-0321
estimated by \citet{b7} was $\sim 7.4 \times 10^{14} \, h_{100}^{-1}
\, \rmn{M_\odot}$ within a region of a radius $r_{200} = 1.5 \,
h_{100}^{-1} \,\rmn{Mpc}$.

The {\it ROSAT HRI\/} observation of MS 1054-0321 was re-analysed by
\citet{b15} and  new {\it HRI\/} data were included in order to gain
better insight into the main components of MS 1054-0321. These authors
found evidence for a significant clump to the west of the main
component of the cluster. The {\it HRI\/} image was fitted to 1D
spherical and 2D elliptical $\beta$-models. Excluding the
western component, Neumann \& Arnaud found the best fit to the
1D spherical model to have $\beta =0.96_{-0.22}^{+0.48}$ and $r_c =
442_{-104}^{+184} \,h_{50}^{-1}\,\rmn{kpc}$. However, if the western
substructure is included, the best spherical model fit gives $\beta =
2$ and $r_c=840 \, h_{50}^{-1} \,\rmn{kpc}$ with large errors. On the
other hand, an elliptical $\beta$-model fits the main cluster component with
$\beta = 0.73 \pm 0.18$, $r_{c1} = (422\pm 109) \, h_{50}^{-1} \,
\rmn{kpc}$ and $r_{c2} = (298\pm 65)\, h_{50}^{-1}\, \rmn{kpc}$. Considering only
the main component of MS 1054-0321 and using
the best fit temperature from the {\it ASCA\/} spectrum,
\citet{b15} derived  gas masses inside $1.65 \, h_{50}^{-1} \,
\rmn{Mpc}$ of $(1.9 \pm 0.3) \times 10^{14}\, h_{50}^{-5/2} \, \rmn{M_\odot}$
and $2.5 \times 10^{14}\, h_{50}^{-5/2} \, \rmn{M_\odot}$ from the best fit
parameters of the 1D and 2D beta models, respectively. They estimated total
masses of MS 1054-0321 at $1.65 \, h_{50}^{-1} \, \rmn{Mpc}$ of
$2.0_{-0.4}^{+0.8} \times 10^{15} \, h_{50}^{-5/2}\, \rmn{M_\odot}$ and $1.6
\times 10^{15}\, h_{50}^{-5/2}\, \rmn{M_\odot}$ from the 1D and 2D $\beta$-models,
respectively. The gas mass fraction is then about $0.10 \,
h_{50}^{-3/2}$ or $0.16 \,h_{50}^{-3/2}$, depending on the choice of
model.

\citet{b22} analysed a {\it Chandra\/} ACIS-S observation of MS
1054-0321. They estimated the X-ray temperature for the
entire cluster to be $10.4_{-1.5}^{+1.7} \, \rmn{keV}$, and confirmed
the substructure in the {\it ROSAT HRI\/} data. The eastern clump is almost
coincident with the main component of the cluster, and has an X-ray
temperature of $10.5_{-2.1}^{+3.4} \, \rmn{keV}$. The western clump is more
compact and denser than the eastern clump, with a temperature of
$6.7_{-1.2}^{+1.7}\,\rmn{keV}$. The quoted errors are at $90\%$
confidence ranges for the two-parameter fit. Using the temperature of the whole
cluster, ~\citeauthor{b22} estimated a virial mass of $\sim 6.2_{-1.3}^{+1.6}
\times 10^{14}\, h_{100}^{-1}\, \rmn{M_\odot}$ within $r_{200} = 0.76 \,
h_{100}^{-1} \, \rmn{Mpc}$. The authors also estimated the total mass
using a $\beta$-model (setting $\beta = 1$ and
fitting for the cluster core radius), and obtained a mass of $7.4
\times 10^{14} \, h_{100}^{-1} \, \rmn{M_\odot}$ within $r_{200} $.

An investigation of the temperature structure of MS 1054-0321 using
{\it XMM--Newton\/} \citep{b24} indicated a temperature for the whole
cluster of $7.2_{-0.6}^{+0.7} \, \rmn{keV}$, lower than
any of the temperatures reported previously. The two
significant clumps already seen by the {\it ROSAT\/} and the {\it
  Chandra\/} satellites were also seen in the {\it XMM--Newton\/}
data. As in the result reported by \citet{b22}, the
eastern clump has the higher temperature, $8.1_{-1.2}^{+1.3} \,
\rmn{keV}$, consistent with the integrated cluster temperature, and
the western clump, at $5.6_{-0.6}^{+0.8} \, \rmn{keV}$, is 30$\%$
cooler than the eastern clump. Quoted confidence limits are $90\%$ for
one interesting parameter.

\citet{b25} analysed {\it HST\/} ACS weak-lensing and {\it Chandra\/}
X-ray data of MS 1054-0321. The dark matter structure mainly
consists of three dominant clumps with a few minor satellite
groups. The eastern weak-lensing clump from the mass map is not detected in the
X-ray data, and the two X-ray clumps are displaced on the apparent
merger axis relative to the central and western weak-lensing
substructures, probably due to ram pressure. ~\citeauthor{b25} estimated  the
projected mass to be $(7.14 \pm 0.11) \times 10^{14}\,h_{100}^{-1}\,
\rmn{M_\odot}$ at $r = 1 \, \rmn{Mpc}$, consistent with previous
lensing work \citep{b26,b27}. They also fitted the X-ray surface brightness of MS
1054-0321 to an isothermal 1D $\beta$-model, excluding the core region
($r < 45\arcsec$) from the fit, to find  $\beta = 0.78 \pm 0.08$ and $r_c = (16\pm
15)\arcsec$, corresponding to $(85\pm 80) \,h_{100}^{-1}\,
\rmn{kpc}$ in our adopted cosmology. Using an X-ray
temperature of $8.9_{-0.8}^{+1.0}\,\rmn{keV}$, ~\citeauthor{b25} found the
mass within radius 1 Mpc is $(8.4 \pm 0.14) \times 10^{14} \,
h_{100}^{-1}\, \rmn{M_\odot}$, agreeing with their weak-lensing result.

Using OVRO/BIMA interferometric SZ effect data, \citet{b12} estimated the gas
mass of MS 1054-0321 within a radius of $94\arcsec$ ($501\, h_{100}^{-1}\,
\rmn{kpc}$ in our adopted cosmology) to be $ (3.7 \pm 0.6) \times
10^{13} \, h_{100}^{-2} \, \rmn{M_\odot}$ for a temperature of
$10.4_{-2.0}^{+5.0} \, \rmn{keV}$. Using the mean gas mass fraction
of a sample of galaxy clusters reported by \citet{b11}, the
derived total mass was $(4.6\pm0.8)\times 10^{14}\,
h_{100}^{-1}\, \rmn{M_{\odot}}$ within the same radius. \citet{b14}
derived the gas mass fraction within $r_{2500}$ for a sample of
massive clusters including MS 1054-0321 from {\it Chandra\/} together
with OVRA/BIMA data using three different models for the plasma
distribution. The first model, the isothermal
$\beta$-model fit to the X-ray data at radius beyond 100 kpc and to
all of the SZ effect data, yielded $f_\rmn{gas}^{\rmn{X\mbox{-}ray}} =
0.144_{-0.013}^{+0.014}$ and $f_\rmn{gas}^{\rmn{SZ}} =
0.153_{-0.029}^{+0.034}$. The second model, the nonisothermal double
$\beta$-model fit to all of the X-ray and SZ effect data, gave
$f_\rmn{gas}^{\rmn{X\mbox{-}ray}}=0.106_{-0.010}^{+0.009}$ and
$f_\rmn{gas}^{\rmn{SZ}}=0.102_{-0.018}^{+0.020}$. The third, the
isothermal $\beta$-model fit to the SZ data with $\beta$ fixed at 0.7,
yielded $f_\rmn{gas}^{\rmn{SZ}}=0.187_{-0.054}^{+0.081}$. Performing
a joint fit analysis of the same X-ray and SZ effect data, \citet{b39}
determined the gas mass fraction of MS 1054-0321 using an
isothermal $\beta$-model excluding the central region (100 kpc)
from the X-ray data. They found the gas mass fraction within $89\arcsec$
($475\,h_{100}^{-1}\,\rmn{kpc}$ in our adopted cosmology) to be
$0.164\pm0.019$.

In Table \ref{properties_literature} we summarize the reported
measurements of the physical properties of MS 1054-0321. The
numerical values of these measurements are converted to the cosmology
adopted by this work. As can be seen, the results are not entirely
consistent.
\begin{table*}
\begin{minipage}{160mm}
\caption{The physical properties of MS 1054-0321 reported in the
  literature.}
\label{properties_literature}
   \begin{tabular}{@{}lccccc@{}}
   \hline
    Study   &  Method   & Radius & $M_{gas}$ & $M_{tot}$ & $f_{gas}$  \\
       &  & $(h_{100}^{-1}\,\rmn{Mpc})$ &
       $(10^{13}\,h_{100}^{-1}\,\rmn{M_{\odot}})$ &
       $(10^{14}\,h_{100}^{-1}\,\rmn{M_{\odot}})$ &    \\
   \hline
      \citet{b7}  & $M$-$T$ relation & $r_{200}=1.50$   & --- &
          7.4\footnote{\label{1st_fnt}Virial mass.} &---   \\
      \citet{b15} & $\beta$-model\footnote{\label{2nd_fnt}Fitted to
        the X-ray data.}  & 1.16  &
      $13.30\pm2.10$ & $14.00_{-2.80}^{+5.60}$ & 0.095   \\
        & Elliptical $\beta$-model\footref{2nd_fnt} & 1.16  & 17.50 & 11.20 &  0.156
        \\
      \citet{b11} & $\beta$-model\footnote{\label{3rd_fnt}Fitted to the SZ effect
            data.} & $r_{500}=0.86$ & --- & --- & $0.053 \pm 0.028$ \\
      \citet{b12} & $\beta$-model\footref{3rd_fnt} &0.50 & $5.29\pm0.86$ &
      $4.60\pm0.80$ & $0.115_{-0.016}^{+0.013}$ \\

      \citet{b22}  & $M$-$T$ relation & $r_{200}=0.76$  & --- &
      6.2\footref{1st_fnt} & ---   \\
           & $\beta$-model\footnote{\label{4th_fnt}Fitted to the X-ray
             data with $\beta$ fixed at 1.00.}&  $r_{200}=0.76$ & ---& 7.4& ---  \\
      \citet{b25} & Weak-lensing &0.7&---& $7.14\pm0.11$ & --- \\
             & $\beta$-model\footnote{\label{5th_fnt}Fitted to
        the X-ray data with excluding the central region.} & 0.7 & ---& $8.40\pm0.14$ & --- \\
      \citet{b14} & $\beta$-model\footnote{\label{6th_fnt}Fitted
        jointly to the X-ray data beyond the central region and to the
        SZ effect data.} & $r_{2500} \approx
      0.18$\footnote{\label{7th_fnt}We estimated the $r_{2500}$ radius
        using parameters from the best-fit model for the SZ effect
        data and the X-ray data outside the central region.} &
      $0.74_{-0.34}^{+0.35}$ & $0.52_{-0.27}^{+0.32}$ &
      $0.144_{-0.013}^{+0.014}$ \\
          &  & $ $ & $0.78_{-0.33}^{+0.29}$ &
          $0.52_{-0.27}^{+0.32}$ &  $0.153_{-0.029}^{+0.034}$   \\
       & Double $\beta$-model\footnote{\label{8th_fnt}Fitted jointly to the
         X-ray and SZ effect data.} & $r_{2500} \approx 0.18$\footref{7th_fnt} & $0.74\pm0.18$ &
       $0.70_{-0.20}^{+0.24}$ & $0.106_{-0.010}^{+0.009}$  \\
         &  & $ $ & $0.72\pm0.11$ & $0.70_{-0.20}^{+0.24}$ &
         $0.102_{-0.018}^{+0.020}$  \\
         & $\beta$-model\footnote{\label{9th_fnt}Fitted to the SZ
           effect data with $\beta$ fixed at 0.70.} & $r_{2500}
         \approx 0.18$\footref{7th_fnt}  & $0.82\pm0.11$ &
         $0.43_{-0.11}^{+0.15}$ & $0.187_{-0.054}^{+0.081}$  \\
       \citet{b39} & $\beta$-model\footref{6th_fnt} & $r_{2500}=0.48$ & $5.18_{-0.70}^{+0.84}$ &
       $3.15_{-0.70}^{+0.98}$ & $0.164\pm0.019$  \\
\hline
\end{tabular}     
\end{minipage}
\end{table*}

\section{DATA}
\subsection{The X-ray image}

\begin{figure*}
  \includegraphics[width=123 mm]{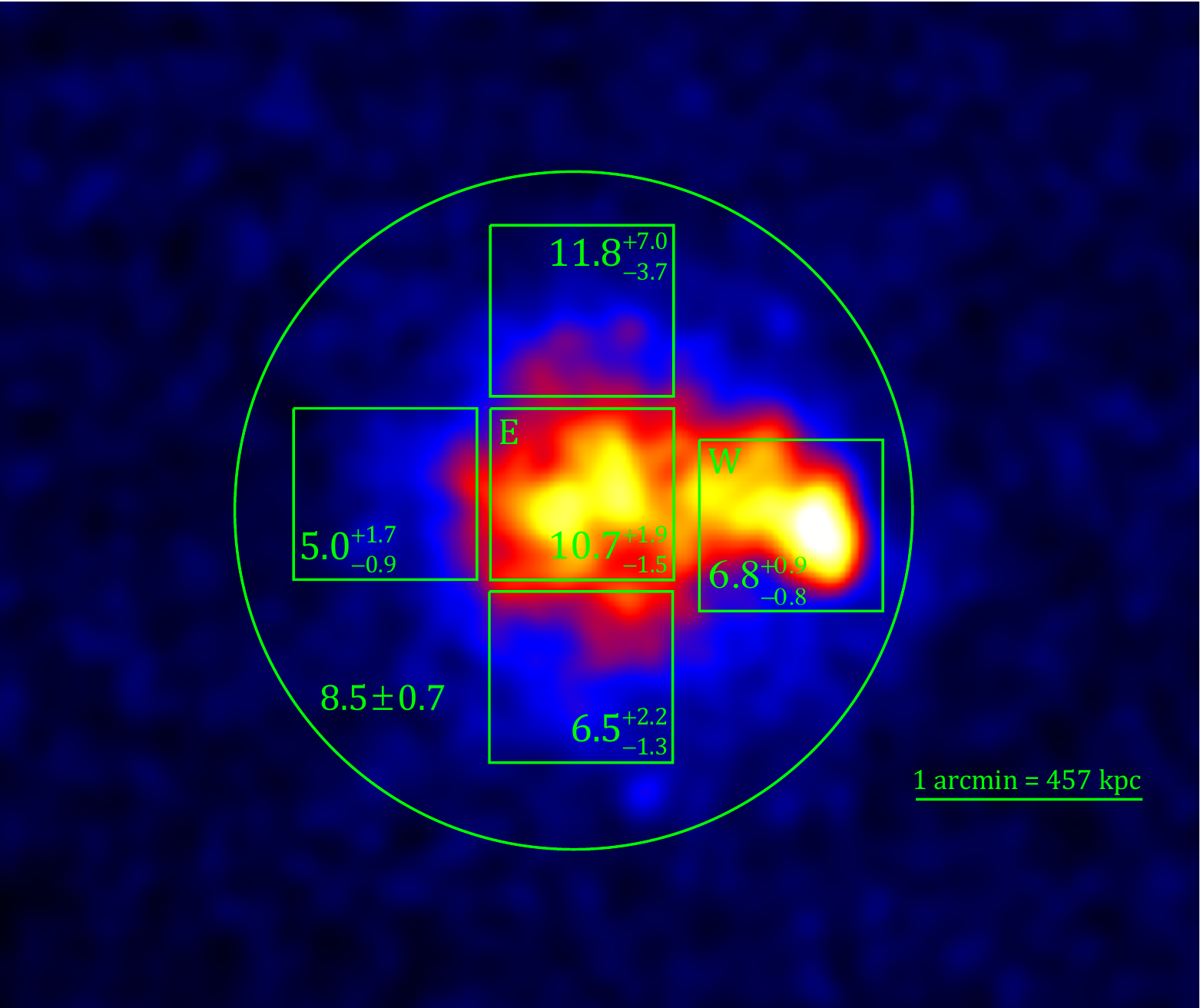}
  \caption{{\it Chandra\/} X-ray image of the cluster MS 1054-0321 in
    the 0.5-7.0 keV energy band smoothed by a Gaussian of $\sigma =
    3.44\arcsec$ (26 kpc). The green circle and boxes show the regions
    in which the X-ray temperatures were extracted. Numerical values are
    the best-fit temperatures in units of keV for given regions. The E
    and W letters denote the eastern and western components
    of the cluster discussed in the main text.}
  \label{ms1054_image}
\end{figure*}

We used archival data for MS 1054-0321 from the same  {\it Chandra\/}
ACIS-S observation in April 2000 used by \citet{b22} and \citet{b25}. The total
observation duration was 90 ks. Chandra Interactive Analysis of
Observations (CIAO) version 4.5, with calibration database (CALDB)
version 4.5.9, was used to reduce the X-ray data. The data were
reprocessed by running  the chandra\_repro script to create a new
level = 2 event file, and a new bad pixel file. To minimize the high
energy particle background, counts were considered only in
the energy range 0.5-7.0 keV. X-ray background counts were
obtained from the local background on the {\it Chandra\/} ACIS-S3
chip. Point sources on this chip  were detected, removed, and replaced
by a local estimate of the background emission. Our image of the X-ray surface
brightness of MS 1054-0321 is shown in Fig.~\ref{ms1054_image}. This
image is smoothed by a two-dimensional Gaussian kernel with
$\sigma=3.44\arcsec$.

The X-ray structure of the cluster shows an elliptical shape
elongated in the east-west direction with two main components,
indicated by two green boxes labelled E and W. The eastern component
is relatively diffuse and less bright at its centre, which lies around
$5\arcsec$ (40 kpc) from the position of the cluster brightest optical
galaxy \citep{b28}. The western component is more compact, has a
higher X-ray surface brightness, and its peak is separated  from the
peak of the eastern component by $\sim 55\arcsec$ (420 kpc).

\subsection{The X-ray spectra}

The emission weighted temperature of the whole cluster was obtained by
using counts ($\sim$ 14120 counts between 0.5 and 7.0 keV) from a
circular region of radius $90\arcsec$ centred at (RA, Dec) = ($10^{\rmn{h}}
57^{\rmn{m}} 00\fs02$, $-03\degr 37\arcmin 36\farcs00$). This
corresponds to the position of the optically brightest
galaxy in the cluster \citep{b28}. We fitted the X-ray spectrum with a
MEKAL model \citep{b29} modified by local Galactic absorption after
binning the counts to at least 30 per energy bin. The iron abundance
and the temperature  were left free to vary, whereas the Galactic
absorption and the cluster redshift were constrained at
$ 3.67 \times 10^{20} \, \rmn{atoms \, cm^{-2}}$ \citep{b30} and 0.83
\citep{b31}, respectively. The best fit yields a temperature $T =
8.5 \pm 0.7 \, \rmn{keV}$ and a metal abundance $Z = 0.33 \pm
0.11 \, Z_\odot$ at 68\% confidence level, with a reduced $\chi ^2 = 0.74$ for 215 degrees of
freedom. Our temperature estimate is in good agreement with the {\it
  Chandra\/} temperature reported by \citet{b25} ($T =
8.9_{-0.8}^{+1.0} \, \rmn{keV}$) and by \citet{b32} ($T =
8.0 \pm 0.5  \, \rmn{keV}$), and also with the {\it XMM--Newton\/}
temperature,  $T = 7.2_{-0.6}^{+0.7} \, \rmn{keV}$ ($90\%$ confidence limits), presented by
\citet{b24}. However, it is slightly lower than the \citet{b22} value,
based on the same {\it Chandra\/} data $(T = 10.4_{-1.5}^{+1.7} \,
\rmn{keV}$ at $90\%$ confidence limits), and much lower than the ASCA
temperature, $T = 12.3_{-2.2}^{+3.1} \,
 \rmn{keV}$ ($90\%$ confidence limits), determined by \citet{b7}. The
 metal abundance obtained from our fitting of the X-ray spectrum
 agrees with previous results reported by \citet{b24} ($Z = 0.33_{-0.18}^{+0.19} \,
 Z_\odot$ at $90\%$ confidence limits) and by \citet{b25} ($Z = 0.30 \pm 0.12 \, Z_\odot$).

\begin{figure*}
  \includegraphics[width=178
 mm]{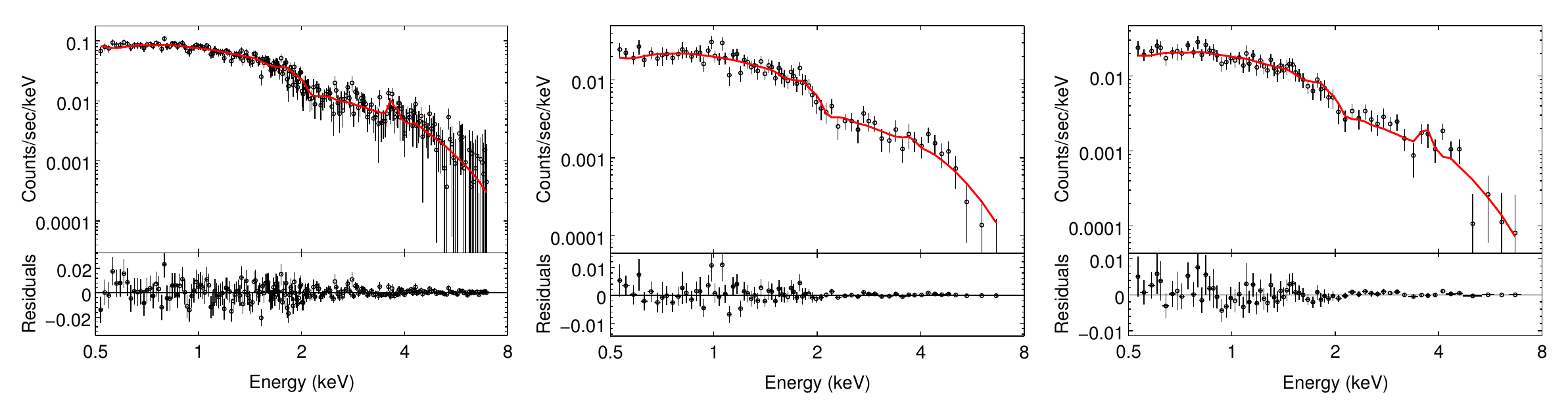}
  \caption{Binned X-ray spectra for MS 1054-0321. Left: The full
    cluster. Centre: the eastern component. Right: the western
    component. The solid line is the best fit thermal plasma model. It
    is clear that the emission line of iron is more evident
    in the wastern component than the eastern component. Lower panels:
    the residuals from the best fit model.}
  \label{ms1054_spec}
\end{figure*}

We also investigated the temperatures of the densest parts of the eastern and western
components by extracting counts from the regions with size of
$48.71\arcsec \times 45.46\arcsec$, centred at RA =
$10^{\rmn{h}}56^{\rmn{m}}59\fs87$, Dec = $-03\degr 37\arcmin
31\farcs68$, and RA = $10^{\rmn{h}}56^{\rmn{m}}56\fs16$, Dec =
$-03\degr 37\arcmin 39\farcs95$ (Fig.~1). There are around 3090 and 2725 counts
in the energy range of 0.5-7.0 keV in the eastern and western
components, respectively. The X-ray data again were fitted with a
MEKAL model, with Galactic absorption frozen at $3.67 \times
10^{20} \, \rmn{atoms \, cm^{-2}}$
and the cluster redshift frozen at 0.83. The temperature and the
metal abundance were left free to vary. The best fit for the
eastern component gives a temperature $ T = 10.7_{-1.5}^{+1.9} \,
\rmn{keV}$ and an abundance $ Z = 0.11_{-0.11}^{+0.22} \, Z_\odot$ with a
reduced $\chi ^2 = 0.71$ for 72 degrees of freedom. For the western
component, the best fit gives a temperature $ T = 6.8_{-0.8}^{+0.9} \,
\rmn{keV}$ and an abundance $ Z = 0.34_{-0.19}^{+0.20} \, Z_\odot$ with a
reduced $\chi ^2 = 0.66$ for 67 degrees of freedom. The X-ray spectra
with the best fit models of the full cluster, the eastern and western components are
shown in Fig.~\ref{ms1054_spec}. All the spectra were
binned to 30 or more counts per energy bin.

Our temperature measurements clearly show that the eastern component is hotter than
the western component. Such results are fully consistent with the
temperatures of the eastern and western components reported by
\citet{b25} ($T = 10.7_{-1.7}^{+2.1}\,\rmn{and}\, 7.5_{-1.2}^{+1.4} \,
\rmn{keV}$), and by \citet{b22} ($T = 10.5_{-2.1}^{+3.4}\,
\rmn{and}\, 6.7_{-1.2}^{+1.7} \, \rmn{keV}$ at $90\%$ confidence
limits). \citet{b24} also noticed the significant difference of
temperature between the eastern and western components ($T =
8.1_{-1.2}^{+1.3}\, \rmn{and}\,
5.6_{-0.6}^{+0.8} \, \rmn{keV}$ at $90\%$ confidence
limits), but their temperatures are systematically lower than our
values. Our results for the abundance agree
with eastern and western component values obtained by \citet{b25}
($Z=0.16_{-0.16}^{+0.19}$ and $0.47_{-0.23}^{+0.24}\, Z_\odot$), and
by \citet{b24} ($ Z=0.12_{-0.12}^{+0.35}$ and $0.51_{-0.32}^{+0.36}\,
Z_\odot$ at $90\%$ confidence limits). \citet{b22} reported similar results,
($Z=0.08_{-0.08}^{+0.23}$ and $0.46_{-0.26}^{+0.27} \,Z_\odot$ at $90\%$ confidence
limits), relative to the slightly different solar iron abundance of
\citet{b33}.

The significant variation in temperature between these two components
led us to investigate the temperature structure for other regions
within the cluster atmosphere. Thus, we extracted counts from three
regions to the north, south, and east of the eastern component (see
Fig.~1). The size of each region was chosen to be equivalent to that of the eastern
(or western) component, and their centres are about $50 \arcsec$ (380
kpc) away from the centre of the eastern component. There are around
1040, 1430, and 890 counts in the energy range 0.5-7.0 keV in the
north, south, and east regions, respectively. The X-ray spectrum in
each region was binned to include at least 50 counts per bin. The
X-ray data were then fitted to a MEKAL model modified by a local
Galactic absorption. Galactic absorption and the cluster redshift were
fixed as before, whereas the temperature and the metal abundance were
left free to vary. The best fit for the region north of the eastern
component gives a temperature $T = 11.8_{-3.7}^{+7.0}\, \rmn{keV}$ and
an abundance $Z = 0.76_{-0.76}^{+1.33} \, Z_\odot$ with a reduced
$\chi^2 = 1.39$ for 15 degrees of freedom. For the region to the
south, the best fit yields a temperature $T = 6.5_{-1.3}^{+2.2}\,
\rmn{keV}$ and an abundance $Z = 0.38_{-0.38}^{+0.46}\,Z_\odot$
with a reduced $\chi^2 = 0.91$ for 22 degrees of freedom. The best fit
temperature and abundance for the region to the east are $T =
5.0_{-0.9}^{+1.7}\, \rmn{keV}$ and $Z = 2.20_{-1.21}^{+2.19}\,
Z_\odot$ with a reduced $\chi^2 = 1.08$ for 13 degrees of freedom.

These results indicate that the temperature of the eastern component
decreases by a factor $\sim 2$ over scale $\sim 50\arcsec$ (380 kpc),
except to the north, where a lesser, or no, temperature change is
found. Fixing the metallicity at the cluster central value leads to
higher outer temperatures in the east and south, and lower temperature
gradients. If the Galactic hydrogen column is allowed to be free, then
we obtain fits of equivalent quality to these found with fixed $N_H$,
but with best-fit temperatures $\sim 1\,\rmn{keV}$ higher in all
cases. However, the fitted values of $N_H$ are below the known
Galactic column, and so we don't consider these results further.

\subsection{The Sunyaev-Zel'dovich effect data}

MS 1054-0321 was observed at 30 GHz over the period 28 February 2006
to 2 June 2006 using the Toru\'n 32 m telescope equipped with the
prototype One Centimetre Receiver Array (OCRA-p) \citep{b34}. OCRA-p
consists of two 1.2 arcmin FWHM beams separated by 3.1 arcmin, and can
achieve 5 mJy sensitivity in 300 seconds. OCRA-p observed MS 1054-0321
for $\sim 11$ hours using a combination of beam-switching and
position-switching. Such a technique is efficient at subtracting
atmospheric and ground signals, and removes such contaminating signals
better than position-switching or beam-switching alone
\citep{b35}. The principles of the beam- and position-switching
technique and the data reduction process are described
in detail in \citet{b34}.

The OCRA data were calibrated by reference to two bright radio sources
NGC7027 and 3C286, with flux densities of 5.64 and 2.51 Jy. By
reprocessing the OCRA data for the MS 1054-0321 cluster, we estimated
the SZ effect toward the cluster centre to be $ -2.18\pm 0.33 \,
\rmn{mJy}$ with standard deviation of individual 1-minute
records $\sigma = 5.16 \, \rmn{mJy}$. This central measured SZ
flux density could be increased or decreased by the emission of radio
sources. An investigation of radio point sources in the MS
1054-0321 field and their impact on the SZ effect measurement is
presented below.

\subsection{The radio environment}

Point source contamination is a major problem in
low-resolution observations of the SZ effect. For the two-beam
receiver, OCRA-p, radio sources could lead to an overestimate of the SZ
effect if they are located in the reference background regions of the
observation. Radio sources lead to an underestimate of the SZ signal
when they lie along the line of the sight to the cluster. By knowing
the flux densities and locations of radio point sources near our
pointing centre at our observing frequency, we can correct for them and
improve our estimate of the SZ effect of the cluster. Currently,
a sensitive survey of the radio sky at 30 GHz does not exist, but the
radio field of the cluster MS 1054-0321 has been observed at multiple
bands. Hence we estimate the flux densities at 30 GHz of
the radio sources in the field of MS 1054-0321 by extrapolating
from lower frequency measurements. We use flux densities from the NRAO
VLA Sky Survey (NVSS) \citep[1.4 GHz,][]{b36}, and the
28.5 GHz flux densities for sources near the cluster centre reported
in \citet{b37}. We also use flux density measurements at 4.85 GHz by
reprocessing archival VLA data. In all fields, only the radio sources
within a radius of 4.5 arcmin from the centre of the cluster are
considered, and we neglect possible radio source variability.

We find four significant radio point sources in the field of the MS 1054-0321
cluster. One is located close to the cluster centre, and another is
found outside the reference arc of the beam-switched observation. The
two remaining sources are found in the reference arc, and they could
produce a negative signal that boosts the apparent SZ effect. All four
point sources have 1.4 GHz flux density measurements. Measurements at
28.5 GHz are available only for the inner three sources.

The archival VLA data were taken at 4835 and 4885 MHz with 50 MHz
bandwidth. We reduced the data using the Astronomical Image Processing
System (AIPS). The source 3C286 = 1331+305 was used as the primary
calibrator to calibrate the flux density, with flux densities of 7.51
and 7.46 Jy. Source 1058+015 was the phase calibrator, and was
observed at around 30 minute intervals throughout the
observation. After the normal calibration and imaging steps, the
position, peak intensity, and total flux density of the detected radio
sources were estimated using the IMFIT task, after correction for the
VLA primary beam.

We estimated the flux densities of the radio sources at 30 GHz by
fitting a power law to the measurements and extrapolating to 30
GHz. The flux densities and positions of the four radio
sources within 4.5 arcmin of the pointing centre of the MS 1054-0321
cluster are given in Table \ref{radio_sources}. These values are
essentially the same as those given by \citet{b37} at 28.5 GHz, except
for source 4 which is not in their list.
\begin{table*}
\begin{minipage}{129mm}
\centering
\caption{The radio source fluxes within 4.5 arcmin from the
     pointing centre of the cluster MS 1054-0321. From left to right the
     columns give the source identification number, the right
     ascension and declination of the radio source, the flux density
     at 1.4 GHz (from~\citealt{b36}), the flux density at 4.8 GHz after primary beam
     correction, the flux density at 28.5 GHz (from~\citealt{b37}), the expected flux
     density at 30 GHz.}
\label{radio_sources}
   \begin{tabular}{@{}lcccccc@{}}
   \hline
    Source ID     &   RA     & Dec & $S_{1.4}$ & $S_{4.85}$ & $S_{28.5}$  & $S_{30}$ \\
       & (J2000) & (J2000) & (mJy) & (mJy) & (mJy) & (mJy)  \\
   \hline
      1  &10 56 59.59  & -03 37 27.7  & $14.10\pm 0.90$ & $7.28\pm 0.10$ \
         & $0.94\pm 0.06$ & $0.87\pm 0.06$  \\
      2  &10 56 57.94  & -03 38 58.3  & $3.10\pm 0.40$ & $1.29\pm
      0.12$ & $0.54\pm 0.08$ & $0.48\pm 0.07$ \\
      3  &10 56 48.85  & -03 37 28.8  & $18.20\pm 1.00$ & $4.41\pm
      0.20$ & $1.79\pm 0.19$ & $1.75\pm 0.19$ \\
      4  &10 56 48.59  & -03 40 07.1  & $4.64\pm 0.40$ & $2.79\pm
      0.22$ & --- & $1.36\pm 0.31$ \\
\hline
\end{tabular}
\end{minipage}
\end{table*}

\subsection{Radio source corrections}

We correct our central SZ effect measurement using the flux densities
and positions of the point sources. The greatest contamination comes from
source 3, which is observed in the reference arc at parallactic angle
$\sim 5\degr$. Source 1 is un-modulated by the changing parallactic
angle during beam-switching, while source 4 has little effect since
it lies far from the sampled reference arc. The corrected value of the
central SZ effect flux density of MS 1054-0321 is $-2.55 \pm 0.33 \,
\rmn{mJy}$, indicating an effect about $1\sigma$ greater than our uncorrected
value, and consistent with a massive hot atmosphere.

\section{MODELING}
\subsection{Cluster atmosphere}

The X-ray surface brightness, $ S_X$, and the Sunyaev-Zel'dovich
effect, $\Delta T$, of a cluster can be expressed as
functions of the electron density, $n_e$, and electron temperature,
$T_e$, of the intracluster plasma integrated along the line of sight
\begin{equation}
S_X={{1}\over{4 \pi(1+z)^4}}D_A\int n_e^2\Lambda_{ee} d\theta_z,
\end{equation}
and
\begin{equation}
\Delta T=f(x) T_r {{k_B \sigma_T}\over{m_ec^2}}D_A\int n_e T_e d\theta_z.
\end{equation}
$D_A$ is the angular diameter distance, $z$ (=0.83) is
the cluster redshift, $\Lambda_{ee}$ is the X-ray spectral emissivity of
the cluster plasma, $f(x)$ describes the frequency dependence of the
SZ effect and has value  $-1.91$ at our observing frequency. $T_r$
\citep[= 2.725 K,][]{b23} is the temperature of the cosmic microwave
background radiation, $k_B$ is the Boltzmann constant, $\sigma_T$ is
the Thomson scattering cross section, $m_e$ is the electron mass, $c$
is the speed of light, and the integration is taken over the line of
sight expressed in angular terms, $d\theta_z$. In order to take into
account the cluster plasma distribution in both components of the MS
1054-0321 cluster, we develop an offset-centred version of a
three-dimensional double $\beta$-model. A centred double $\beta$-model
is often used to fit the X-ray surface brightness for clusters that
show centrally-enhanced density profiles. One component describes the
cluster-centred density peak, while the second component is flatter
and describes the outer part of the cluster. \citet{b3} and
\citet{b14} used a spherical double $\beta$-model to describe density
profiles of cool core clusters. In this work we use a double
$\beta$-model to fit the eastern and western components of the MS
1054-0321 cluster which are centred at positions separated by $\sim
55\arcsec$.

Our three dimensional double $\beta$-model of the electron density is
of the form
\begin{eqnarray}
n_e=n_{e1}+n_{e2} \hspace{14.05 em}\nonumber \\
    =n_{e01}\Big(1+{{\theta_{x1}^2}\over{\theta_{cx1}^2}}+{{\theta_{y1}^2}\over{\theta_{cy1}^2}}+{{(\theta_z-\delta)^2}\over{\theta_{cz1}^2}}\Big)^{-3\beta_1/2}\nonumber\\
 +\,
 n_{e02}\Big(1+{{\theta_{x2}^2}\over{\theta_{cx2}^2}}+{{\theta_{y2}^2}\over{\theta_{cy2}^2}}+{{\theta_z^2}\over{\theta_{cz2}^2}}\Big)^{-3\beta_2/2},\hspace{2.0em}
\label{plasma_model}
\end{eqnarray}
and
\begin{equation}
   \left(
   \begin{array}{c}
   \theta_{xn} \\
   \theta_{yn}
   \end{array}
   \right)=\left(
   \begin{array}{ccc}
   \cos \alpha_n   & \sin \alpha_n   \\
   -\sin \alpha_n  & \cos \alpha_n
   \end{array}
   \right) \left(
   \begin{array}{c}
   \Theta_x-\Theta_{x0n} \\
   \Theta_y-\Theta_{y0n}
   \end{array}
   \right),
\end{equation}
where $n$ takes values of 1 and 2, corresponding to
the eastern or western component. $n_{e01}$ and $n_{e02}$ are the
central electron densities of the eastern and western components,
$\beta_1$ and $\beta_2$ describe the shapes of the
plasma distributions of the eastern and western components,
($\theta_{cx1}$, $\theta_{cy1}$, $\theta_{cz1}$) and ($\theta_{cx2}$, $\theta_{cy2}$, $\theta_{cz2}$)
are the angular core radii of the eastern and western
components. $\delta$ is the offset between the two components
  along the line of sight, $\alpha_n$ describes the rotation angle of
  the $xy$-plane of each component relative to an observer frame,
  ($\Theta_x$, $\Theta_y$) are angular coordinates, and
  ($\Theta_{x0n}$, $\Theta_{y0n}$) are the component centre
  coordinates.

This is not a general triaxial $\beta$-type model, but is adequate to
describe the X-ray and Sunyaev-Zel'dovich surface brightnesses of the
MS 1054-0321 cluster. Under the assumption of an isothermal
intracluster medium, with the electron temperature equal to the
central temperature ($T_{en}=T_{e0n}$), the X-ray spectral emissivity
becomes a constant over each component
($\Lambda_{ee}=\Lambda_{een}$). For non-overlapping atmospheres, the
angular structures of the X-ray surface brightness and
Sunyaev-Zel'dovich effect profile for each component can be expressed as
\begin{eqnarray}
S_{Xn} \propto \sqrt{\pi}{{\Gamma
    (3\beta_n-{{1}\over{2}})}\over{\Gamma({3\beta_n})}}\theta_{czn}
\hspace{10em} \nonumber \\
 \times \Big(1+{{\theta_{xn}^2}\over{\theta_{cxn}^2}}+{{\theta_{yn}^2}\over{\theta_{cyn}^2}}\Big)^{(1/2)-3\beta_n},
\end{eqnarray}
and
\begin{eqnarray}
\Delta T_n \propto
\sqrt{\pi}{{\Gamma({{3}\over{2}}\beta_n-{{1}\over{2}})}\over{\Gamma({{{3}\over{2}}\beta_n})}}\theta_{czn}
\hspace{10em} \nonumber \\
\times \Big(1+{{\theta_{xn}^2}\over{\theta_{cxn}^2}}+{{\theta_{yn}^2}\over{\theta_{cyn}^2}}\Big)^{(1/2)-(3/2)\beta_n},
\end{eqnarray}
respectively. Any overlap between the atmospheres of the two
components will produce a further X-ray emission contribution
$\propto\int n_{e1}n_{e2}d\theta_z$.

Accordingly, if we can approximate the structure in terms of two
distinct components (i.e., if the clusters are separated significantly
beyond the largest core radius), the X-ray and SZ effect central
densities for each cluster component can be obtained using
\begin{equation}
n_{e0n}^{\rmn{X\mbox{-}ray}}=\Big[{{S_{X0n}4\pi(1+z)^4}\over{\Lambda_{een}\sqrt{\pi}
  D_A\theta_{czn}}}{{\Gamma({3\beta_n})}\over{\Gamma({3\beta_n-{{1}\over{2}}})}}\Big]^{1/2},
\end{equation}
and
\begin{equation}
n_{e0n}^{\rmn{SZ}}={{\Delta T_{0n}m_ec^2}\over{T_r f(x) k_B T_{e0n}\sigma_T\sqrt{\pi}D_A\theta_{czn}}}{{\Gamma({{{3}\over{2}}\beta_n})}\over{\Gamma({{{3}\over{2}}\beta_n-{{1}\over{2}}})}},
\end{equation}
respectively. In these equations $S_{X0n}$ and $\Delta T_{0n}$ are the
central X-ray surface brightness and SZ effect signal of each cluster component.

\subsection{Cluster gas and total masses}

The total number of electrons in the cluster can be calculated by
integrating the electron density profile over a given volume. The
cluster gas mass can be obtained by multiplying the total number of
electrons by the mean mass per electron, $\mu_e m_p$, so that the gas mass
for each component of the cluster in some volume takes the form
\begin{eqnarray}
M_{gas}(\theta_{xn},\theta_{yn},\theta_z)=\mu_e m_p D_A^3
\iiint n_{en} d\theta_{xn} d\theta_{yn} d\theta_z.
\end{eqnarray}
The total mass of the cluster, which is dominated by dark matter,
can be deduced from the cluster gas structure. Its distribution
relates to the electron density and temperature profiles of the
plasma. We assume that the cluster gas is in hydrostatic equilibrium
in the gravitational potential well of the cluster with no significant
flow of matter. With the further assumption of an isothermal plasma, the
total mass in each cluster component
\begin{eqnarray}
M_{tot}(\theta_{xn},\theta_{yn},\theta_z)={{-k_BT_{e0n}}\over{4\pi G \mu
    m_p}}D_A\iiint\Big({{\partial^2}\over{\partial\theta_{xn}^2}}+{{\partial^2}
  \over{\partial\theta_{yn}^2}}\nonumber
\\+{{\partial^2}\over{\partial\theta_z^2}}\Big) \ln n_{en} d\theta_{xn}d\theta_{yn}d\theta_z ,
\label{total_mass}
\end{eqnarray}
where $G$ is the Newtonian gravitational constant, and $\mu$ is the
mean mass per particle in unit of $m_p$.

\subsection{Fit model to the X-ray image}

The X-ray surface brightness image of the MS 1054-0321 cluster in the
0.5-7.0 keV energy band was fitted to our 3D $\beta$-model. In
order to use the $\chi^2$ statistic on X-ray event data, we re-binned
the data so that every bin contains at least 20 counts. Below we
summarise the data gridding and fitting processes, and their
implications.

First we identified the location of each event in the 2D surface
brightness image. We divided the 2D image vertically into slices with
equal widths, and then we divided each slice horizontally into bins in
such a way that every bin contains at least 20 counts. This mean that
bins have areas with equal width but different height based on the
density of events in a given location of the 2D image. In addition to
calculating the number of counts in each bin, we also calculated the
location of the count centroid and the bin area. As a result of the
binning process, the X-ray data roughly became normally distributed,
and this allows us to apply the $\chi^2$ statistic and deal with fewer
data events, therefore making the fitting process faster, without
losing much detail of features of the X-ray surface brightness image.

Typically, the X-ray surface brightness images contain some regions
with few counts. For the MS 1054-0321 cluster, we noticed some bins
with very small values of the  number of counts per unit area. In
order to get a better fit to the X-ray data, we excluded bins with
values smaller than 0.03 counts pixel$^{-1}$ from our data
analysis. We chose to use Lmfit version 0.7.4 as an optimizer and the
Levenberg-Marquardt algorithm as the optimization method in
Python.\footnote{See http://lmfit.github.io/lmfit-py.} We measure the
goodness of the fit using the $\chi^2$ statistic, in this case given
as
\begin{equation}
\chi^2=\sum\limits_i\Big({{D_i-M_i}\over{\epsilon_i}}\Big)^2,
\end{equation}
where $D_i$ is the number of counts detected per unit area in
bin $i$, $M_i$ is the number of counts predicted by the model per
unit area in bin $i$, and $\epsilon_i$ is Poisson error on $D_i$.

We first fit a single $\beta$-model plus a background component to the
eastern component of the cluster, i.e. excluding the X-ray data of the
western component and setting  $n_{e02}=0$ in equation
(\ref{plasma_model}). Because the X-ray data cannot constrain all the
parameters in the 3D model, we took the
core radius along the line of sight, $\theta_{cz1}$, to equal
$\theta_{cx1}$ or $\theta_{cy1}$. The centre of the eastern component
was fixed to the position of the brightest cluster galaxy,
whereas the other parameters were left free to vary. Values of the
best fit parameters for the single $\beta$-model are shown in Table
\ref{single_beta}. We present only the results with the assumption of
$\theta_{cz1}=\theta_{cy1}$ since it provided  slightly the better
fit. The value of the reduced $\chi^2$ associated with this fit is 1.40
with 1041 degrees of freedom. Hence this fit can be rejected at the
99.9$\%$ confidence level.
\begin{table}
\begin{minipage}{70mm}
\centering
\caption{Three-dimensional single $\beta$-model fit.}
\label{single_beta}
\begin{tabular}{@{}l c c c@{}}
\hline
Parameter         &  Fitted value\footnote{Best-fit values with $68\%$
  error bounds.}         &   &       Unit     \\
\hline
$S_{X01}$          &  $(6.01 \pm 0.06)\times 10^{-2}$ &  &     counts
s$^{-1}$ arcmin$^{-2}$    \\
$\alpha_1$       &   $6.76  \pm 0.07$          & &degrees     \\
$\theta_{cx1}$        &  $0.90 \pm 0.08$   &  &    arcmin         \\
$\theta_{cz1}=\theta_{cy1}$    &   $1.09 \pm  0.05$   &  &    arcmin         \\
$\beta_1$          & $1.26 \pm 0.09$        &   & ---             \\
$C$\footnote{The local X-ray background.} & $(6.22 \pm 0.08)\times 10^{-3}$  &  & counts s$^{-1}$ arcmin$^{-2}$    \\
\hline
\end{tabular}
\end{minipage}
\end{table}

A better fit for the X-ray surface brightness image of MS 1054-0321 is
provided by the 3D double $\beta$-model. Here, the
  X-ray data were fitted to the model with $\delta$ set to three
  different values: 0, 1, and 3 Mpc. The component centres and other
parameters in equation (\ref{plasma_model}) were left free to
vary. The fit qualities are nearly equivalent in all cases,
  but a slightly better fit is  obtained with $\delta = 0$. For both
  components, the optimised values obtained for the
  central surface brightness from all fits differ by less than
  $3\%$. The shape parameters ($\beta, \theta_c$) of the eastern and
  western components at $\delta = 1\,\rmn{Mpc}$ differ from those
  obtained at $\delta = 3\,\rmn{Mpc}$, and from those obtained at our
  adopted offset $\delta = 0$, but the differences do not exceed the
  errors.

\begin{figure*}
  \includegraphics[width=155 mm]{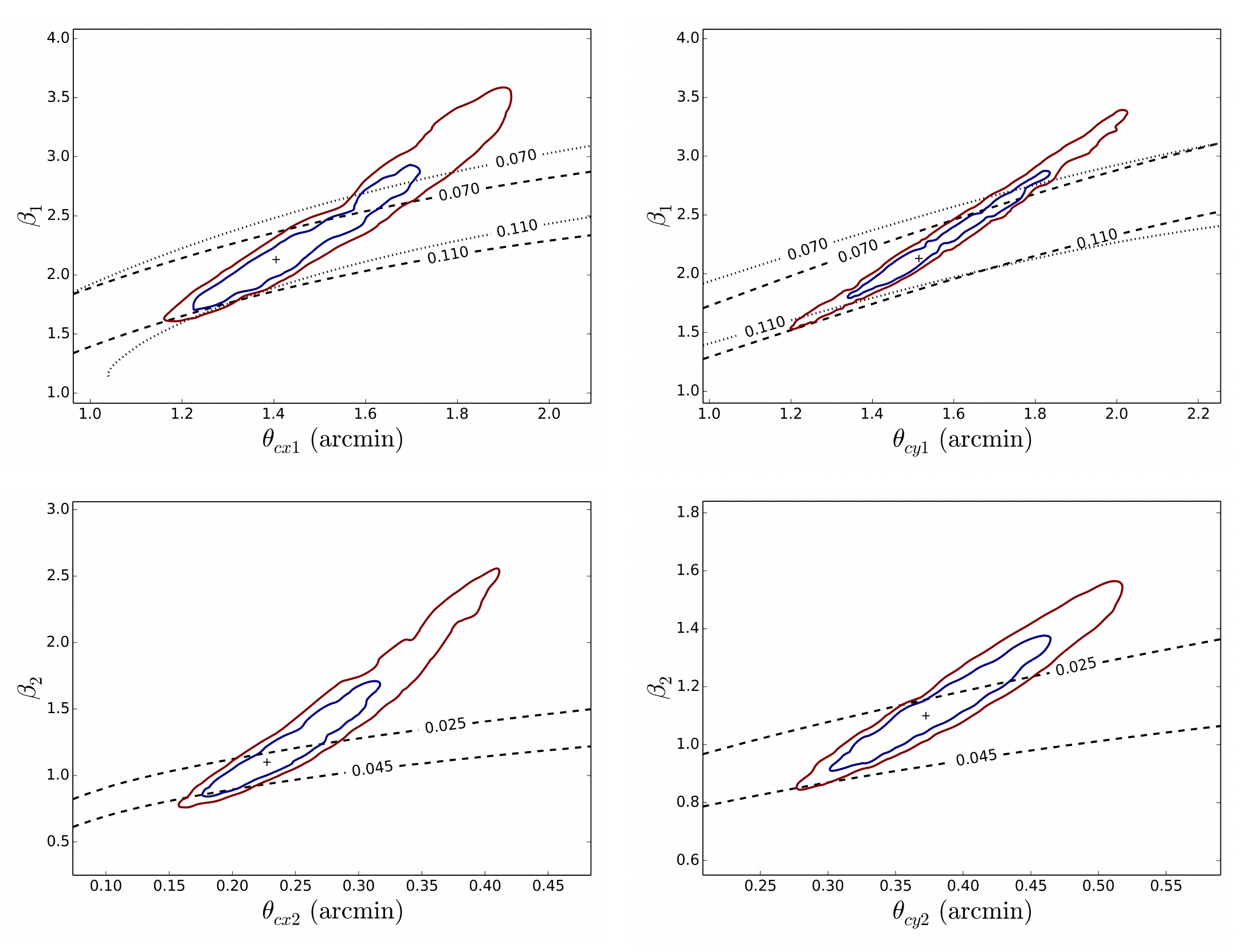}
  \caption{Confidence interval contours for the double $\beta$-model
    fit to the X-ray data of the MS 1054-0321 cluster, superposed with
    contours of the values of the gas mass fractions from the
    X-ray and SZ effect data. The best fit of the parameters are
    marked as crosses, while the blue and red solid contours correspond,
    respectively, the $68\%$ and $90\%$ confidence intervals.  The
    dashed and dotted contours define, respectively, the gas mass
    fractions derived from the X-ray and SZ effect data.Upper
    and lower panels represent the confidence level contours of the
    shape parameters for the eastern and western components,
    respectively.}
  \label{correlation_ms1054}
\end{figure*}

\begin{figure*}
  \includegraphics[width=155 mm]{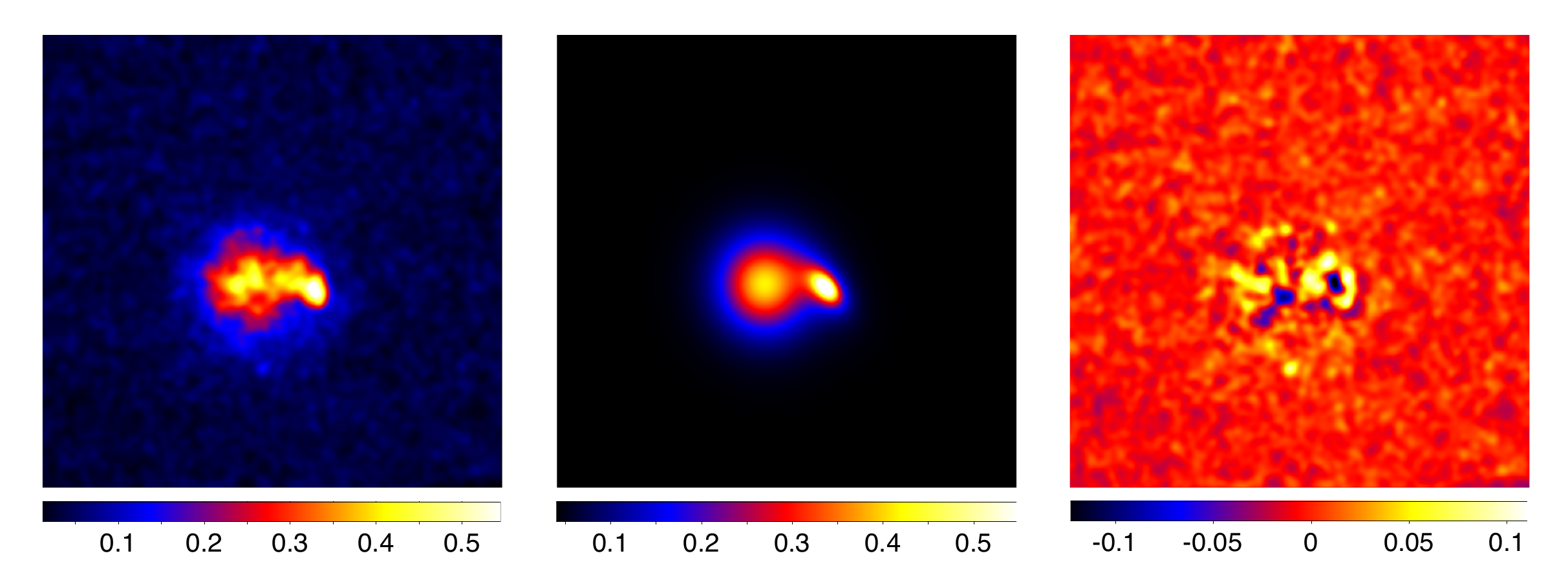}
  \caption{Left: {\it Chandra\/} X-ray image of the cluster MS 1054-0321 in the
    0.5-7.0 keV band. Centre: the source model, made up of two
    elliptical $\beta$ models. Right: the residual image from the
    best-fit model. The colour scales at the bottom of each
    image are in units of counts pixel$^{-1}$, where a pixel has side
    $0.492\arcsec$.}
  \label{data_model_resid}
\end{figure*}

The best fit parameters and single parameter $68\%$ errors for the
 double $\beta$-model in case of $\delta = 0$ are given in Table
 \ref{double_beta}. As before, we present only the results with
the assumptions of $\theta_{cz1}=\theta_{cy1}$ and
$\theta_{cz2}=\theta_{cy2}$. We found the reduced $\chi^2$ associated
with this fit is 1.06 with 1423 degrees of freedom. Clearly, the X-ray
data are better described by this model than the
single $\beta$-model. The estimated centre of the
eastern component is less than $6\arcsec$ away from the position of the
brightest galaxy in the cluster, and the centre of the western component is
consistent with the position of the brightest pixel in the
X-ray image. The best fit value of the shape parameter obtained from this
fit is particularly high for the eastern component, $\beta_1=2.1 \pm
 0.4$ and a value of $\beta_1<1$ is excluded at high significance
($\Delta\chi^2> 39$). This characteristic has previously been
reported~\citep[see][for
example]{b15,b22,b25}. Fig.~\ref{correlation_ms1054} shows the
confidence intervals for key shape parameters of the eastern and
western components. This figure shows the usual strong correlation
between $\beta$ and $\theta_c$. Fig.~\ref{data_model_resid}
illustrates the X-ray surface brightness of the cluster MS 1054-0321,
the best double $\beta$-model fit to the X-ray data, and the residual
image from the best fit.
\begin{table}
\begin{minipage}{70mm}
\centering
\caption{Three-dimensional double $\beta$-model fit with $\delta = 0$.}
\label{double_beta}
\begin{tabular}{@{}l c c c@{}}
\hline
Parameter       &  Fitted value\footnote{Best-fit values with $68\%$
  error bounds.}     &   &       Unit     \\
\hline
$S_{X01}$        &    $(6.22 \pm 0.17)\times 10^{-2}$ &   & counts s$^{-1}$ arcmin$^{-2}$    \\
$\Theta_{x01}$        & 10 56 59.64 $\pm$ 0.08     & &hr, min, sec  \\
$\Theta_{y01}$         & -03 37 36.75 $\pm$ 0.56     & & deg, arcmin, arcsec  \\
$\alpha_1$       &   $12.62 \pm 0.39$          & &degrees     \\
$\theta_{cx1}$          &  $1.41\pm 0.19$    &  &  arcmin         \\
$\theta_{cz1}=\theta_{cy1}$    &   $1.51\pm 0.21$ &   &    arcmin         \\
$\beta_1$         &  $2.13\pm 0.43$        & & ---             \\
$S_{X02}$          &  $(10.85 \pm 0.98)\times 10^{-2}$ &  &     counts s$^{-1}$ arcmin$^{-2}$    \\
$\Theta_{x02}$        &  10 56 55.82 $\pm$ 0.08     & &hr, min, sec  \\
$\Theta_{y02}$         & -03 37 39.72 $\pm$ 0.70     & &deg, arcmin, arcsec  \\
$\alpha_2$       &   $40.03 \pm 2.89$          & &degrees     \\
$\theta_{cx2}$          & $0.23\pm 0.05$      &   &arcmin         \\
$\theta_{cz2}=\theta_{cy2}$      &  $0.38\pm 0.08$  &  &    arcmin         \\
$\beta_2$           & $1.10\pm 0.24$         &  & ---             \\
$C$\footnote{The local X-ray background.}   & $(6.24 \pm
0.07)\times 10^{-3}$  & &counts s$^{-1}$ arcmin$^{-2}$    \\
\hline
\end{tabular}
\end{minipage}
\end{table}
\section{RESULTS AND DISCUSSION}
\subsection{Central electron density}

For our single $\beta$-model fit, the estimated
central densities of the eastern component are
$4.96_{-0.11}^{+0.09}\times 10^{-3} \, \rmn{cm}^{-3}$ and
$5.06_{-0.06}^{+0.09}\times 10^{-3} \, \rmn{cm}^{-3}$ from the X-ray
and SZ effect data, respectively. For the double $\beta$-model
  with zero offset between the two components along the line of sight,
  the corresponding values are $5.00_{-0.03}^{+0.01}\times 10^{-3} \,
  \rmn{cm}^{-3}$ and $5.35_{-0.15}^{+0.01}\times 10^{-3} \,
  \rmn{cm}^{-3}$. The estimated X-ray central density of the western
component is $10.38_{-0.02}^{+0.39}\times 10^{-3}\,\rmn{cm}^{-3}$. SZ
effect data are not available for the western component. For the
eastern and western components, the derived central densities from the
X-ray and SZ effect data at the 1 and 3 Mpc offsets differ by less
than about $3\%$ from those derived at $\delta = 0$.

As clearly indicated by these results, the X-ray central densities
obtained for the eastern component using the single and double
$\beta$-models are consistent with each other, and with the central
density values derived from the SZ effect data. They are also in good
agreement with those found from a nonisothermal centred double
$\beta$-model \citep{b14}. The central density of the western
component is more than twice that of the eastern component.
\subsection{Gas mass fraction}

The gas mass fraction, $f_{gas}$, can be computed by dividing the
gas mass by the total mass of the cluster. To determine the cluster
masses as discussed in section 4.2, we need to define a limiting
radius within which the masses are measured. Due to
significant AGN activity in MS 1054-0321 relative to lower redshift
clusters, particularly at distances between 1 and 2 Mpc
\citep{b28,b38}, we carried out our measurements within the relatively
small angular radius $\theta_{2500}$, corresponding to a density
2500 times the critical density of the Universe at the
redshift of the cluster. The $\theta_{2500}$ radius of each cluster
component can be obtained from
\begin{equation}
M_{tot}(\theta_{2500})=D_A^3\theta_{2500}^3\Delta \rho_c(z),
\end{equation}
where the left-hand side is the total mass within angular radius
($\theta_{2500}$)  given by equation (\ref{total_mass}), $\rho_c(z)$ is
the critical density of the Universe, and $\Delta = 2500$.

In Table \ref{gas_fraction}, we show the gas and total masses, and the
gas mass fractions, of the eastern and western components using the
best fit parameters from our models with $\delta = 0$. The 68$\%$
confidence interval errors associated with these are given for
variations of the fitting parameters. For both models, the gas mass
fractions derived from the X-ray and SZ effect data for the eastern
component are consistent. It can be seen from the double $\beta$-model
analysis that the X-ray gas mass fraction of the western component
within $\theta_{2500}$ is about one third of that derived for the
eastern component.
\begin{table*}
\begin{minipage}{150mm}
\caption{The physical properties of the MS 1054-0321 cluster.}
\label{gas_fraction}
   \begin{tabular}{@{}lcccccc@{}}
   \hline
   \multicolumn{7}{ |c| }{Single $\beta$-model} \\
   \hline
   Region & $\theta_{2500}$ & $M_{gas}^\rmn{X\mbox{-}ray}$ & $M_{gas}^\rmn{SZ}$ &
   $M_{tot}$& $f_{gas}^\rmn{X\mbox{-}ray}$ & $f_{gas}^\rmn{SZ}$ \\
         & (arcsec) & $(10^{13}\, h_{100}^{-1}\, \rmn{M_{\odot}})$ &
         $(10^{13} \, h_{100}^{-1}\,\rmn{M_{\odot}})$ & $(10^{14}\,
         h_{100}^{-1}\, \rmn{M_{\odot}})$& & \\
   \hline
   Eastern component  & $40.9_{-1.0}^{+0.3}$  & $1.30_{-0.01}^{+0.02}$ &
   $1.33_{-0.01}^{+0.04}$ & $1.45_{-0.02}^{+0.01}$ &
   $0.090_{-0.001}^{+0.001}$ & $0.092_{-0.001}^{+0.001}$ \\
   \hline
   \multicolumn{7}{ |c| }{Double $\beta$-model} \\
   \hline
   Eastern component & $44.8_{-1.0}^{+3.5}$ & $1.67_{-0.01}^{+0.06}$ &
   $1.79_{-0.05}^{+0.06}$ &  $1.91_{-0.11}^{+0.01}$ &
   $0.087_{-0.001}^{+0.005}$ &  $0.094_{-0.001}^{+0.003}$ \\
   Western component & $43.9_{-4.9}^{+10.5}$ &  $0.53_{-0.12}^{+0.04}$ & --- &
   $1.79_{-0.36}^{+0.82}$ &  $0.030_{-0.014}^{+0.010}$ & ---  \\
   \hline
\end{tabular}
\end{minipage}
\end{table*}

Although the values of the shape parameters extracted from the models
are significantly different, we find that the gas and
total masses, and thus the gas mass fraction, for the eastern component
obtained from the single $\beta$-model are consistent with those
derived from the double $\beta$-model over the same angular
radius. The X-ray and SZ effect gas mass fractions
from both models are in good agreement with those derived from a
nonisothermal double $\beta$-model  within $\theta_{2500}$
\citep{b14}, and $\sim 70\%$ higher than that derived by \citet{b11}
within a larger angular radius. However, our gas
mass fractions are $\sim 40\%$ lower than those derived from an
isothermal $\beta$-model fit that excluded the central region from the
X-ray data \citep{b14}.

To compare our results with those derived by \citet{b39}, we estimate the gas and
total masses within their angular radius $\theta =
89\arcsec$. Within this radius, the derived X-ray and SZ effect gas
masses for the eastern component from the single $\beta$-model
analysis are, respectively, $4.40_{-0.01}^{+0.02} \times
10^{13}\,h_{100}^{-1}\,\rmn{M_{\odot}}$ and $4.51_{-0.01}^{+0.06} \times
10^{13}\,h_{100}^{-1}\,\rmn{M_{\odot}}$. From the double $\beta$-model
fit, these gas masses are $4.36_{-0.04}^{+0.20} \times
10^{13}\,h_{100}^{-1}\,\rmn{M_{\odot}}$ and
$4.67_{-0.08}^{+0.16}\times
10^{13}\,h_{100}^{-1}\,\rmn{M_{\odot}}$. These values are consistent with
each other and with the gas mass of $5.18_{-0.70}^{+0.84} \times
10^{13}\,h_{100}^{-1}\,\rmn{M_{\odot}}$
derived by~\citet{b39} within the same angular radius from an isothermal
$\beta$-model fit to the X-ray data (excluding the central
region) and to SZ effect data. Using the
temperature of the eastern component, $10.7_{-1.5}^{+1.9}\,
\rmn{keV}$, our estimates of the total mass inside 89$\arcsec$ from
the single and double $\beta$-models are $5.64_{-0.43}^{+0.68} \times
10^{14}\,h_{100}^{-1}\,\rmn{M_{\odot}}$ and $7.70_{-1.00}^{+0.91} \times 10^{14}\,h_{100}^{-1}\,\rmn{M_{\odot}}$, respectively. Such
masses are $\sim 60\%$ larger than the mass
of $3.15_{-0.70}^{+0.98} \times
10^{14}\,h_{100}^{-1}\,\rmn{M_{\odot}}$ given by \citet{b39}. The
estimated total masses fall to $4.48_{-0.34}^{+0.54} \times
10^{14}\,h_{100}^{-1}\,\rmn{M_{\odot}}$ and $6.12_{-0.80}^{+0.73} \times 10^{14}\,
h_{100}^{-1}\,\rmn{M_{\odot}}$ from the single and double
$\beta$-models if we adopt the lower temperature of $8.5 \pm
0.7\,\rmn{keV}$ measured for the total cluster. The former value is
more consistent with that reported by \citet{b39} than the total mass
derived from the double $\beta$-model. The high total mass
  that we derived from the double $\beta$-model, relative to that
  derived from the single $\beta$-model, might suggest that the total
  mass is overestimated as a result of overlaps between the two
  components. The gas mass fractions  obtained within $89 \arcsec$
for the eastern component are
$f_{gas}^{\rmn{X\mbox{-}ray}}=0.102_{-0.011}^{+0.009}$ and
$f_{gas}^{\rmn{SZ}}=0.127_{-0.013}^{+0.010}$ from the single
$\beta$-model analysis, and
$f_{gas}^{\rmn{X\mbox{-} ray}}=0.074_{-0.006}^{+0.015}$ and
$f_{gas}^{\rmn{SZ}}=0.096_{-0.007}^{+0.017}$ from the double
$\beta$-model analysis. The average value of these gas mass fractions
is $0.100_{-0.009}^{+0.013}$, about $ 40\%$ lower than the gas mass
fraction estimated by \citet{b39}.

For the eastern component, the estimated X-ray and SZ gas mass
fractions within $\theta_{2500}$ using the 1 Mpc offset parameters are
consistent with those estimated using the 3 Mpc offset parameters, and
only about $5\%$ higher than those derived at $\delta = 0$. These
results imply an systematic error of $\sim 0.002$ in the X-ray and SZ
gas mass fractions due to uncertainty in the value of $\delta$. The
X-ray gas mass fractions for the western component derived within
$\theta_{2500}$ for 1 and 3 Mpc offsets agree, but are about $10\%$
higher than the value for $\delta = 0$, suggesting an systematic error
of $\sim 0.001$ in the gas mass fraction. These results indicate that
the offset issue is not responsible for the low gas mass fraction in
the eastern and western components.

The gas mass fraction of the eastern and  western components together,
within angular radius $\theta_{2500}$, is
$0.060_{-0.009}^{+0.004}$. This fraction is small, and
about two thirds of the gas mass fraction deduced from the eastern
component. However, the value is in agreement with that reported
by \citet{b11} within $\theta_{500}$, and $\sim 40\%$ lower than the X-ray and SZ effect
gas mass fractions measured from the nonisothermal double $\beta$-model \citep{b14}.

To examine the effect of the isothermal assumption on the gas mass
fraction measurements, we estimated the gas and total masses for both
cluster components in the presence of the maximum temperature gradient
(Section 3.2). For the eastern component, the estimated
X-ray and SZ gas masses are higher by about $3\%$ and $20\%$ than
those derived under the isothermal assumption, respectively. The total
mass could be as much as $50\%$ lower than that derived
from the isothermal assumption. The upper limits on the gas mass
fractions are $\sim 0.140$ from the X-ray data, and $\sim 0.180$ from
the SZ effect data. The estimated X-ray and SZ gas mass fractions
therefore are
$f_{gas}^{\rmn{X\mbox{-}ray}}=0.087_{-0.001}^{+0.005}\,\rm{(random)}\,{}_{-0.001}^{+0.053}\,\rm{(systematic)}$
and
$f_{gas}^{\rmn{SZ}}=0.094_{-0.001}^{+0.003}\,\rm{(random)}\,{}_{-0.001}^{+0.086}\,\rm{(systematic)}$.
For the western component, the X-ray gas mass fraction is affected by
less than $10\%$ by the isothermal assumption. Thus the gas mass
fraction in the eastern component could approach the cosmic value,
while the western component is gas-poor. Any contribution from the
non-thermal pressure would tend to reinforce our conclusion of low gas
mass fraction.

Such a low gas mass fraction could suggest
that the gas content of the cluster, particularly the western component,
is being stripped. This scenario was also proposed by
\citet{b22}. Gas stripping in this cluster is likely since it contains
a large fraction of merging galaxies \citep[17$\%$,][]{b31}, which
might indicate that MS 1054-0321 has not yet reached a steady
state. Accordingly, measurements of the component gas mass fractions
under the assumptions of isothermal and hydrodynamic equilibrium
conditions could be questionable. The stronger dependence of the SZ
gas mass on temperature suggest that a sensitive SZ map extending to
larger radii from the cluster centre might reveal the missing gas mass
if the outer gas is relatively cool.

\section{Summary}

Using archival {\it Chandra\/} data, we
have re-analysed the temperature structure of the MS 1054-0321
cluster. The best fit model to the cluster X-ray spectrum within
$90\arcsec$ of the cluster centre gives
X-ray temperature $T=8.5 \pm 0.7\,\rmn{keV}$ and iron abundance
$Z=0.33\pm0.11\, \rmn{Z_{\odot}}$. The measured temperature is
consistent with the {\it Chandra\/} temperature reported by
\citet{b25} and \citet{b32}, and also with the {\it XMM--Newton\/}
temperature \citep{b24}. However, this temperature is lower by $\sim
18\%$ than the {\it Chandra\/} temperature derived by \citet{b22}, and
by $30\%$ than the ASCA temperature \citep{b7}. The iron abundance is
in agreement with the values reported by \citet{b24} and
\citet{b25}. The temperature of the eastern component of MS
  1054-0321 is significantly higher than that of the western component
  (Section 3.2). Such a temperature difference is consistent with
previous work \citep{b22,b24,b25}.

We have measured the central SZ effect signal at 30 GHz towards MS
1054-0321 with OCRA-p. The contaminated SZ effect signal towards the
cluster centre is $-2.18 \pm 0.33\, \rmn{mJy}$. The radio
environment within an angular radius of 4.5 arcmin of the cluster
centre was investigated, and used to form a source-corrected SZ effect of
$-2.55 \pm 0.33\, \rmn{mJy}$.

We have investigated the gas mass distribution of
the MS 1054-0321 cluster using triaxial $\beta$-type models. In
addition to fitting the X-ray data of the eastern
component with a single triaxial $\beta$-model, we have fitted the X-ray
image of the entire cluster to a double triaxial $\beta$-model, which
gives a significantly better description of the gas structure of the
cluster. For the single and double $\beta$-models, the gas mass
fractions estimated from the X-ray and SZ effect data for the eastern
component are in good agreement. Using different offsets along the
line of sight, results obtained from the double $\beta$-model suggest
that the gas mass fraction of the western component is about three
times lower than those of the eastern component.

Despite a significant variation in the shape parameters for the
models, the X-ray and SZ gas mass fractions for the eastern component
derived from the single $\beta$-model are consistent with those
derived from the double $\beta$-model within the same region.
For both models, the gas mass fractions are in good agreement with
those derived from a nonisothermal $\beta$-model \citep{b14}, and
higher than the gas mass fraction reported by \citet{b11} over
the larger region. The X-ray and SZ gas mass within angular
radius $89\arcsec$ from both models agree well with those
reported by \citet{b39}. Using the integrated X-ray
  temperature of the cluster, the average gas mass fraction estimated
  for the eastern component at $\delta = 0$ is
  $0.100_{-0.009}^{+0.013}$, about $40\%$ lower than that reported by
  \citet{b39}. The gas mass fraction of the overall cluster, within
$\theta_{2500}$, is $0.060_{-0.009}^{+0.004}$. This is in agreement
with the gas mass fraction reported by \citet{b11}. The relative
model-independence of these gas mass fractions suggests that these
estimates are robust and reliable.

We have quantified the effect of the isothermal assumption and
hydrostatic equilibrium condition on measurements of the gas and total
masses, and thus the gas mass fraction of the cluster. For the eastern
component, the X-ray and SZ gas mass fractions could rise to $\sim
0.140$ and $\sim 0.180$, respectively. For the western component, the
X-ray gas mass fraction is less affected by the isothermal
assumption. Contributions from non-thermal pressure could lower the
estimated gas content of this component even further.

We suggest that the low gas mass fraction of the cluster,
particularly in the western component, could be a result of gas
stripping. This scenario is consistent with the merger picture of the
cluster \citep{b31}, and the assumptions of the isothermal and
hydrostatic equilibrium conditions are therefore
questionable. Observations with better signal/noise in the SZ effect
could reveal more about the nature of this cluster and locate the
missing baryons, and significantly deeper {\it Chandra\/} observations
could resolve uncertainties about the temperature structure of the
eastern component.

\label{lastpage}
\end{document}